\documentclass{article}

\usepackage{PRIMEarxiv}
\usepackage{amssymb}
\usepackage[T1]{fontenc}
\usepackage{graphicx}
\usepackage{color}
\graphicspath{{images/}} 
\usepackage{color, soul}
\usepackage{microtype}
\soulregister\cite7
\soulregister\ref7
\soulregister\pageref7
\usepackage{multirow}
\usepackage{comment}
\usepackage{amssymb}
\usepackage{pifont}
\usepackage{amsmath} 
\usepackage{dsfont}
\newcommand{\cmark}{\ding{51}}%
\usepackage{tikz}
\usetikzlibrary{shapes.geometric, arrows}
\tikzstyle{box} = [rectangle, rounded corners, minimum width=0.2\linewidth, minimum height=1cm,text centered, text width=0.2\linewidth, draw=black, fill=white!30]

\tikzstyle{arrow} = [thick,->,>=stealth]

\title{How Secure is Forgetting?\\ Linking Machine Unlearning to Machine Learning Attacks}

\author{
Muhammed Shafi K. P. \\
Department of Computer Applications, \\
Cochin University of Science \\
and Technology, India \\
\texttt{shafikp@cusat.ac.in} \\
\And
Serena Nicolazzo \\
Department of Computer Science, \\
University of Milan, \\
G. Celoria, 20, Milan, Italy\\
\texttt{serena.nicolazzo@unimi.it} \\
\And
Antonino Nocera \\
Department of Electrical, Computer \\
and Biomedical Engineering, \\
University of Pavia, \\
A. Ferrata, 5, Pavia, Italy \\
\texttt{antonino.nocera@unipv.it} \\
\And
Vinod P. \\
Department of Computer Applications, \\
Cochin University of Science \\
and Technology, India \\
\texttt{vinod.p@cusat.ac.in} \\
}

\begin{document}

\maketitle

\date{March 2025}

\begin{abstract}
As Machine Learning (ML) evolves, the complexity and sophistication of security threats against this paradigm continue to grow as well, threatening data privacy and model integrity. In response, Machine Unlearning (MU) is a recent technology that aims to remove the influence of specific data from a trained model, enabling compliance with privacy regulations and user requests. This can be done for privacy compliance (e.g., GDPR's right to be forgotten) or model refinement. 
However, the intersection between classical threats in ML and MU remains largely unexplored. In this Systematization of Knowledge (SoK), we provide a structured analysis of security threats in ML and their implications for MU. We analyze four major attack classes, namely, Backdoor Attacks, Membership Inference Attacks (MIA), Adversarial Attacks, and Inversion Attacks, we investigate their impact on MU and propose a novel classification based on how they are usually used in this context. Finally, we identify open challenges, including ethical considerations, and explore promising future research directions, paving the way for future research in secure and privacy-preserving Machine Unlearning.

\end{abstract}


\section{Introduction}

The term Machine Unlearning (MU, hereafter) identifies the process of removing the influence of specific data points from a trained Machine Learning (ML) model while preserving model performance and efficiency. This can be done for several reasons, namely privacy compliance, reducing biased data, or redundant training samples. As for privacy compliance, one of the most significant provisions introduced by regulations such as the General Data Protection Regulation (GDPR) in the European Union and the California Consumer Privacy Act (CCPA) in the US is the Right to Be Forgotten (RTBF), which allows individuals to request the deletion of their personal information from databases, online platforms, and digital records \cite{rosen2011right}. In the ML context, the RTBF requires that both the data and their influence on the model must be removed \cite{bourtoule2021machine}.
Other reasons for employing the MU technique are: {\em(i)} removing mislabeled, outdated, or biased data in the computation to improve model fairness and {\em (ii)} deleting redundant training samples to reduce model complexity.
A straightforward approach for deleting unwanted data from the ML computation would be to retrain ML models from scratch while excluding the data that have to be unlearned. However, this method is highly computationally expensive and often impractical for real-world applications \cite{bourtoule2021machine}. For this reason, researchers are investigating new techniques that do not include full retraining of ML models but attempt to just update them. These models have been increasingly deployed in critical applications that require high levels of trust and reliability. 
Furthermore, while MU strengthens data control and regulatory compliance, the consequences of classical ML security and privacy vulnerabilities, including adversarial exploitation, data leakage risks, and model integrity threats, in such systems remain largely unexplored and should be carefully analyzed to ensure trustworthiness \cite{li2023reconstructive,chen2021machine}. 
In light of these challenges, an exhaustive examination of existing research on the implications of ML threats and MU systems is essential.
In this Systematization of Knowledge (SoK), we comprehensively analyze the classical attacks on security and privacy in ML and their intersection with MU systems. We provide a novel perspective that focuses on the categorization of the literature according to the main known attacks and how they are used in the context of MU. Although some recent survey contributions have begun to examine MU approaches, the majority of existing works focus on MU fundamental concepts and methodologies \cite{li2025machine,shaik2024exploring,xu2024machine,sai2024machine}. Other contributions focus on the evolution of MU within the Federated Learning settings, i.e., Federated Unlearning (FU) that has emerged to confront the challenge of data erasure in distributed learning environments \cite{liu2024survey,romandini2024federated}.
Whereas, Blanco et al. \cite{blanco2025digital} surveys MU methods for LLMs. By contrast, \cite{chen2025survey,liu2025threats} adopt a different perspective; as a matter of fact, the work of Chen et al. \cite{chen2025survey} focuses on privacy risks associated with the adoption of MU exploring existing countermeasures for model protection from malicious unlearning-based attacks, whereas \cite{liu2025threats} deals only with an analysis of current threats and defenses in MU. 

To the best of our knowledge, the SoK proposed in this paper is the first to offer a comprehensive perspective and a thorough examination of the key vulnerabilities of ML and their relationship with MU solutions. 
Indeed, at first, we have selected relevant existing studies discussing MU together with ML attacks. Then, by examining these contributions, we identified the most used types of attack against ML, namely Backdoor Attacks, Membership Inference Attacks, Adversarial Attacks, and Inversion Attacks. From our analysis, we discovered that four key types of relationships may occur: {\em(i)} a known ML attack can be perpetrated against the MU system, {\em(ii)} MU can be used as a defense mechanism to mitigate an ML attack, {\em(iii)} an ML attack can be employed as evaluation tools to assess the effectiveness of a new MU frameworks, and {\em(iv)} an ML attack can be exploited as tools for evaluating a new MU verification approach. 

In this study, we conducted a comprehensive search for publications related to the aforementioned topics. Specifically, our review focused on journal and conference papers published in the past years, sourced from Google Scholar, Web of Science, ACM Digital Library, IEEE Xplore, USENIX, and SpringerLink. Additionally, we examined Grey Literature, including white papers and government reports. Our selection criteria prioritized Q1 and Q2 journals, A-ranked conference papers, and high-quality white papers from reputable sources such as government agencies, industry leaders, and academic institutions. We excluded duplicate entries and non-English papers. We analyzed a total of 65 works selecting 32 papers from 2020 to 2025. 

Our systematic categorization aims to identify existing gaps in the interplay between attacks on ML and MU systems, encouraging researchers to improve both the robustness and reliability of current defense strategies and find new solutions. The main contributions of this paper are the following.
\begin{itemize}
    \item We present a solid description of Machine Unlearning including the recent techniques and evaluation metrics used in this context.
    \item We introduce a taxonomy that systematically categorizes existing attacks in the context of ML and their implications in Machine Unlearning solutions.
    \item We present a comprehensive analysis of the challenges associated with security threats in MU. In addition, we highlight critical areas for future research and development in this field.
\end{itemize}

The outline of this paper is as follows. In Section \ref{sec:mul}, we present the essential background of MU necessary to understand our paper. In Section \ref{sec:attacks}, we examine the papers dealing with ML attacks and MU systems and propose a novel classification for them. Section \ref{sec:challenge} examines present challenges and limitations that can lead to future work. Finally, Section \ref{sec:conclusion} draws our conclusions.

\section{Background on Machine Unlearning}
\label{sec:mul}
This section provides the essential background information to understand the key concepts discussed in this paper. In particular, it offers a concise overview of Machine Unlearning categories, key techniques, and the common metrics used to evaluate approaches in this context.

Table \ref{tab:SystemSymbols} summarizes the acronyms used in this paper.

\begin{table}
\centering
  \caption{Summary of the acronyms used in the paper}
  \begin{tabular}{ll}
\hline
    \textbf{Symbol} & \textbf{Description}\\
\hline
    ATMs & Adversarial Training Models\\
    BA & Backdoor Attack\\
    CMU & Centralized Machine Unlearning\\
    DNNs & Deep Neural Networks\\
    DL & Deep Learning\\
    FL & Federated Learning\\
    FU & Federated Unlearning\\
    GIA & Gradient Inversion Attack\\
    IA & Inversion Attack\\
    MIA & Membership Inference Attack\\
    ML & Machine Learning\\
    MLaaS & ML-as-a-Service\\
    MoIA & Model Inversion Attack\\
    MU & Machine Unlearning\\
\hline
\end{tabular}
\label{tab:SystemSymbols}
\end{table}

\subsection{Definition and Classification}

\begin{figure}[ht]
    \centering
    \includegraphics[scale=0.7]{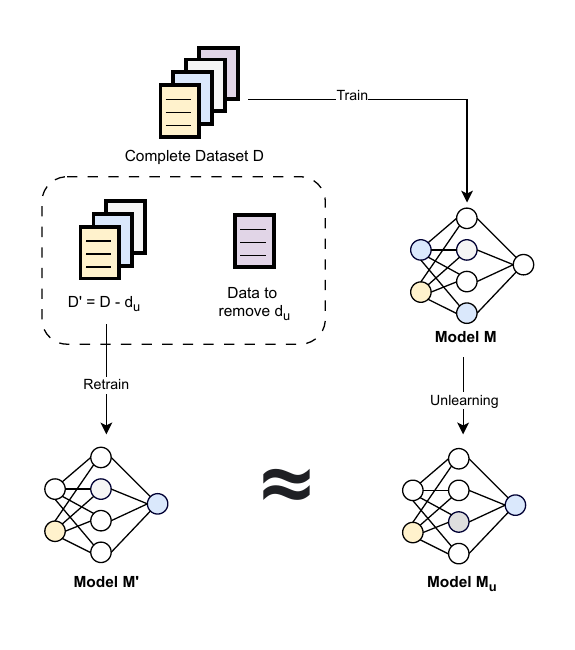}
    \caption{Machine Unlearning workflow}
    \label{fig:MU}
\end{figure}

As stated in the Introduction, Machine Unlearning (MU) refers to the process by which a system removes the influence of previously learned data that was incorporated through an ML algorithm \cite{cao2015towards}. Ideally, the model completely forgets the data points that contributed to its training, effectively eliminating their impact, so that if the same data point is reintroduced in the future, the system processes it as if it were entirely new, without any residual knowledge from prior learning. In formulas,
let $D=\{(x_1,y_1), \cdots, (x_n,y_n)\}$ represent the original and complete training dataset, where $x_i$ is the input feature and $y_i$ the corresponding label and $M$ is the ML model trained on $D$. Let $d_u$ represent a set of data points to be removed so that $D'\subseteq D$ and $ D' = D - d_u$, where $D'$ is the dataset excluding the removed data. When the unlearning starts all information related to $D_u$ should be deleted from the trained model $M$. This is obtained by constructing an unlearned model denoted as $M_u$ that should be indistinguishable from the retrained model $M'$ and obtained through retraining from scratch data in $D'$.
Figure \ref{fig:MU} visually represents the MU process of removing specific data points from a trained ML model while preserving overall functionality.

MU techniques can be classified according to the exactness of the obtained unlearning in the two following typologies \cite{sai2024machine,li2025machine}:
\begin{itemize}
    \item \textbf{Exact Unlearning.} Perfect or exact unlearning algorithms aim to produce a model that is identical to one trained from scratch on a dataset excluding a specific data point that needs to be unlearned. This represents the ideal scenario, but achieving such precision is highly challenging. As a result, retraining the model from the ground up is currently considered the only true exact unlearning method.
    \item \textbf{Approximate Unlearning.} Approximate unlearning offers a more cost-effective alternative and is particularly beneficial for complex and adaptive ML algorithms, where reconstructing the precise sequence and impact of individual data points is often infeasible. 
\end{itemize}

An additional MU classification relies on the level of guarantee that the specified data points have been completely erased from the model. Based on this difference we can distinguish the following types of MU methods:

\begin{itemize}
    \item \textbf{Certified MU} that gives the formal guarantee that a model from which data is removed cannot be distinguished from a model that never observed the data to begin with \cite{guo2023certified}. Even if this method gives strong privacy and security assurance, retraining is resource-intensive and may not be feasible for large-scale models.
    \item \textbf{Empirical MU} does not ensure strong theoretical guarantees for security and its validation method is based on experimental testing and observation. for these reasons, this kind of MU is used more in real-world applications.
\end{itemize}

Another possible classification is based on how data influence is removed \cite{sai2024machine}. The most common type of removal is item or instance-based and targets the removal of specific data points from a trained model \cite{bourtoule2021machine}. This is the case used for privacy compliance. There might be a case in which all data points that belong to a specific category or class should be removed. For instance in face recognition applications, each face is treated as a separate class. Therefore, when a user decides to stop using the application and requests the removal of their facial data, it falls under this category. Finally, also feature-level MU can be performed, eliminating the impact of specific features while retaining others. This last case is particularly costly because a model has to be developed to recognize and delete the feature usually involving a large amount of data \cite{warnecke2023machine}.

A final popular classification may be conducted according to the adopted paradigm. In this case, we can distinguish between a Centralized MU (CMU), where Unlearning is performed on a single model stored in a central location, and a Federated Unlearning (FU). FU allows the FL model to remove the impact of a worker or of the information associated with a worker's local data while guaranteeing the privacy of the decentralized learning process. This new kind of paradigm introduces new targets and peculiar challenges \cite{liu2024survey}.

\subsection{Key Techniques}
A trivial technique for MU is {\em Retraining from Scratch} in which the model is retrained from the beginning on the dataset excluding the data that needs to be forgotten. Although it is computationally expensive and impractical for large-scale models it provides an exact and certified method for unlearning.
An improved approach that also can be categorized as an exact method is the {\em Sharded or Partitioned Training} that divides the data into disjoint fragments and trains a model on each smaller data fragment. In this way, the training cost can be distributed \cite{bourtoule2021machine}.

Apart from the above-mentioned exact methods, there exist approximated techniques that modify the trained model without full retraining, achieving efficient but non-guaranteed unlearning. The following approaches can be listed among these approximated methods \cite{li2025machine}:

\begin{itemize}
    \item \textbf{Influence Function-Based Unlearning} are based on {\em influence functions} that can quantify the influence of data on learning models, thus estimating the model changes caused by erasing the unusable data. Through these methods, efficient and harmless unlearning can be achieved by updating the trained model according to the estimated changes \cite{zhang2024recommendation}.
    \item \textbf{Knowledge Distillation-Based Unlearning} is based on the concept of Knowledge Distillation that allows the training of a student model (unlearned model) by selectively replicating the knowledge of a larger teacher model (original model). This process enables the removal of sensitive information related to the data to be deleted while preserving the overall utility and effectiveness of the student model. Recent work in~\cite{lee2024learning} alleviated Electric Vehicle user data exposure risks by employing a teacher-student framework that trains a teacher model on the full dataset and a student model to forget sensitive data, using a dual-term loss function to preserve performance and maintain closeness between the unlearned and trained models.

    \item \textbf{Gradient-based Unlearning} approximates the retrained model by correcting the stochastic gradient descent (SGD) steps. It leverages the gradients (i.e., parameter updates) computed during training to estimate and reverse the impact of the data to be forgotten.
    \item \textbf{Federated Unlearning (FU)} allows the selective removal of data influence from a Federated Learning (FL) model without requiring global retraining from scratch.
    FU can be executed {\em(i)} server-side, if the server plays the crucial role of unlearning the global model and redistributing it to the workers, or {\em(ii)} local-side if clients unlearn the downloaded global model before reuploading it to the server \cite{liu2024survey}.
\end{itemize}

\subsection{Evaluation Metrics}

Evaluation metrics allow model providers to measure the effectiveness, utility, and efficiency of their unlearning processes, facilitating the optimization of unlearning algorithms for improved performance. The most employed metrics are the following:

\begin{itemize}
    \item \textbf{Forgetting Rate (FR)} measures the reduction in model performance on the removed data. It can be computed as:
    $$FR = 1- \frac{A_{after}}{A_{before}}$$
    where $A_{after}$ is the accuracy on removed data after Unlearning; and $A_{before}$ is the accuracy on removed data before Unlearning \cite{ma2022learn}.
    \item \textbf{Model Distance Metrics} compares the difference between the unlearned model and a model retrained from scratch without the removed data.
    \item \textbf{Membership Inference Resistance}. Since Membership inference attacks identify a given data sample in the training dataset, if an attack still recognizes the unlearned data as a member after the unlearning, it means that the unlearning process has failed \cite{li2023reconstructive}.
    \item \textbf{Accuracy Drop (AD)} measures the difference in accuracy before and after unlearning the remaining data. It quantifies the degradation introduced by the forget set on the unlearned model and it is computed through the formula:
    $$AD = \frac{|A_{before}-A_{after}|}{A_{before}}$$
    where $A_{after}$ is the accuracy on removed data after Unlearning; and $A_{before}$ is the accuracy on removed data before Unlearning.
    \item \textbf{Attack Success Rate (ASR)} measures the effectiveness of attacks (e.g., adversarial unlearning, re-insertion attacks) against the MU system. In particular, for a Backdoor Attack, it quantifies the ratio of poisoned data that are misclassified into the target label desired by the attacker \cite{liu2022backdoor}. 
    \item \textbf{Unlearning Time (UT)} measures the time (expressed in seconds or number of epochs) required to forget specific data points.
\end{itemize}

\section{Machine Unlearning and Security Attacks}
\label{sec:attacks}
This section systematically explores the most popular security threats in ML and their relationship with Machine Unlearning. We examine all the threats used in the papers we analyzed, namely Backdoor Attacks, Membership Inference Attacks, Adversarial Attacks, and Inversion Attacks. Moreover, we use a categorization highlighting the relationship between these attacks and Machine Unlearning according to what we found in the scientific literature.
Table \ref{tab:papers} lists all the analyzed works according to the described attacks (i.e., Backdoor Attack (BA), Membership Inference Attack (MIA), Adversarial Attack (AA), and Inversion Attack (IA)). In addition, we provide the scope of the paper. In particular, we identified five possible main scopes for the analyzed documents, namely: {\em(i)} they define a new MU method; {\em(ii)} they propose a variation of a known ML attack; {\em(iii)} they describe a possible defense against a known ML attack leveraging MU; {\em(iv)} they design a new metrics to evaluate a MU method; {\em(v)} they define a new MU verification method.

\begin{table*}

\centering
\caption{Analysed papers dealing with ML attack and MU\label{tab:papers}}

\renewcommand{\arraystretch}{1.5}

\begin{tabular}{|l|c|cccc|p{1.2cm}p{1.2cm}p{1.2cm}p{1.2cm}p{1.8cm}|}
\hline
    \multirow{2}*{\textbf{Ref.}} & \multirow{2}*{\textbf{Year}} & \multicolumn{4}{c|}{\multirow{1}{*}{\textbf{Attacks}}}&  \multicolumn{5}{c|}{\multirow{1}{*}{\textbf{Paper Scope}}}\\
    \cline{3-11} 
    & &$BA$ & $MIA$ &$AA$ & $IA$& MU Method & Attack & Defense & Evaluation Metric& MU Verification Method\\
    \hline
    Chen et al. \cite{chen2021machine} & 2021 & - & \cmark & -  & - & 
    - & \cmark & \cmark & \cmark & - \\
    
    Golatkar et al. \cite{golatkar2021mixed} & 2021 & - & \cmark & - & - & 
    \cmark & - & - & - & -\\
    
    Graves et al. \cite{graves2021amnesiac} & 2021 & - & \cmark & - & \cmark & 
    \cmark & - & - & - & -\\
    
    Gupta et al. \cite{gupta2021adaptive} & 2021 & - & - & \cmark & - & 
    \cmark & \cmark  & \cmark & - & -\\
    
    Liu et al. \cite{liu2022backdoor} & 2022 & \cmark & - & - &  - & 
    - & - & \cmark & - & -\\
    
    Ma et al. \cite{ma2022learn} & 2022 & \cmark & \cmark & - & - & 
    \cmark & - & - & \cmark & -\\
    
    Marchant et al. \cite{marchant2022hard}  & 2022 & \cmark & - & -  & - & 
    - & \cmark & - & - & -\\
    
    Sommer et al. \cite{sommer2022athena} & 2022 & \cmark & \cmark & - & - & 
    - & - & - & - & \cmark\\
    
    
    Chundawat et al. \cite{chundawat2023zero} & 2023 & - & \cmark & -  & \cmark & 
    \cmark & - & - & \cmark & -\\
    
    Guo et al. \cite{guo2023verifying} & 2023 & \cmark & - & -  & - & 
    - & - & - & - & \cmark\\
    
    Jia et al. \cite{jia2023model} & 2023 & \cmark  & - & -  & - & 
    \cmark & - & - & - & -\\
    
    Kurmanji et al. \cite{kurmanji2023towards} & 2023 & -  & \cmark & - & - & 
    \cmark & - & - & - & -\\
    
    Liu et al. \cite{liu2023muter} & 2023 & -  & - & \cmark  & - & 
    \cmark & - & \cmark & - & - \\    
    
    Li et al. \cite{li2023reconstructive} & 2023 & \cmark  &  - & - & - &  
    \cmark & \cmark & \cmark & - & -\\ 
    
    Wei et al. \cite{wei2023shared} & 2023 & \cmark  & - &  - & - &  
    - & - & \cmark & - & -\\ 

    Zhang et al. \cite{zhang2023fedrecovery} & 2023 & - & \cmark &  - & - & 
    \cmark & - & - & - & -\\
    
    Zhao et al. \cite{zhao2023static} & 2023 & - & - & \cmark &  - & 
    - & \cmark & - & - & -\\
    
    Daluwatta et al. \cite{daluwatta2024uaas} & 2024 & \cmark & -  & - & - & 
    - & - & \cmark & - & -\\
    
    Chen et al. \cite{chen2024privacy} & 2024 & - & \cmark &  - & - & 
    \cmark & - & - & - & -\\
    
    Chen et al. \cite{chen2024private} & 2024 & \cmark & \cmark &  - & - & 
    \cmark & - & - & \cmark & -\\

    Huang et al. \cite{huang2024uba} & 2024 & \cmark & - & - & - & 
    - & \cmark & - & - & -\\
    
    Gao et al. \cite{gao2024defending}& 2024 & -  & - &  - & \cmark  & 
    - & - & \cmark & - & -\\
    
    Hu et al. \cite{hu2024learn} & 2024 & -  & - & -  & \cmark & 
    - & \cmark & \cmark & - & -\\
    
    Jiang et al. \cite{jiang2024efficient}& 2024& \cmark & \cmark  & - & - & 
    \cmark & - & - & - & -\\

    Liu et al. \cite{liu2024backdoor} & 2024 & \cmark & -  & - & - & 
    - & \cmark & - & - & -\\ 
    
    Li et al. \cite{li2024partially} & 2024 & \cmark & - & -  & - & 
    - & - & \cmark & - & -\\
        
    Zhao et al. \cite{zhao2024facilitating} & 2024 & \cmark & - & -  & - & 
    - & - & \cmark & - & -\\
    
    Niu et al. \cite{niu2024towards} & 2024 & \cmark & - & \cmark & - & 
    - & - & \cmark & - & -\\
    
    Wu et al. \cite{wu2024unlearning}& 2024 &  \cmark & - & - & - & 
    - & - & \cmark & - & -\\
    
    Chen et al. \cite{chen2025fedmua} & 2025 & \cmark & - & - & - & 
    - & \cmark & \cmark & - & -\\    
    
    Han et al. \cite{han2025vertical} & 2025 & \cmark & \cmark &  - & - & 
    \cmark & \cmark & - & - & -\\
    
    Varshney and Torra \cite{varshney2025efficient} & 2025 & - & \cmark & - & - & 
    \cmark & - & - & - & -\\
    \hline
\end{tabular}

\footnotesize $BA$: Backdoor Attack; $MIA$: Membership Inference Attack; $AA$: Adversarial Attack; $IA$: Inversion Attacks.
\end{table*}

\begin{figure*}
\centering
\scalebox{.9}{
\begin{tikzpicture}[node distance=0.2\linewidth]
\node (start) [box,fill={rgb:black,1;white,5}] {Attacks Classification};

\node (n1) [box, below of=start,xshift=-0.4\linewidth] {Backdoor Attack 
\cite{marchant2022hard,huang2024uba,liu2024backdoor,chen2025fedmua,liu2022backdoor,li2023reconstructive,wei2023shared,daluwatta2024uaas,zhao2024facilitating,niu2024towards,jia2023model,ma2022learn,han2025vertical,jiang2024efficient,sommer2022athena,guo2023verifying,han2025vertical}};
\node (n2) [box, below of=start,xshift=-0.15\linewidth] {Membership Inference Attack \cite{chen2021machine,sommer2022athena,zhang2023fedrecovery,jiang2024efficient,han2025vertical,varshney2025efficient,golatkar2021mixed,graves2021amnesiac,ma2022learn,chundawat2023zero,kurmanji2023towards,chen2024privacy,chen2024private}};

\node (n3) [box, below of=start, xshift=0.15\linewidth] {Adversarial Attack \cite{zhao2023static,gupta2021adaptive,niu2024towards,liu2023muter}};
\node (n4) [box, below of=start, xshift=0.4\linewidth] {Inversion Attack \cite{hu2024learn,graves2021amnesiac,chundawat2023zero,gao2024defending}};

\draw [arrow] (start) -- (n1);
\draw [arrow] (start) -- (n2);
\draw [arrow] (start) -- (n3);
\draw [arrow] (start) -- (n4);

\end{tikzpicture}}
\caption{Categorization of papers into ML Attacks} \label{fig:attacksClassification}
\end{figure*}
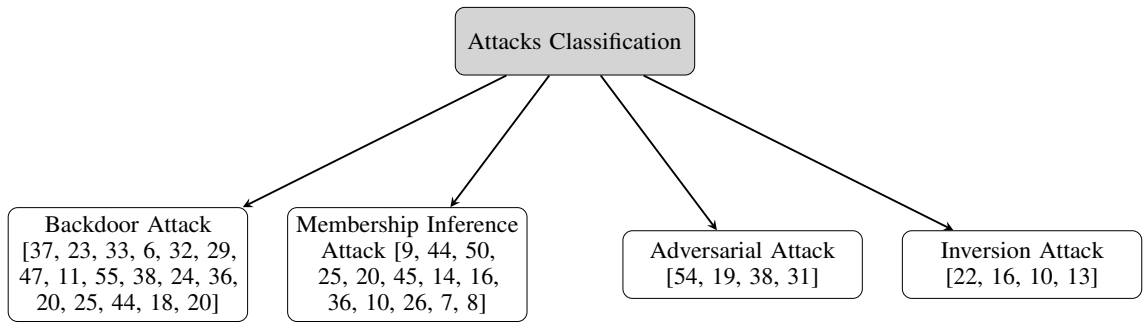

Figure \ref{fig:attacksClassification} represents our adopted classification based on the four known ML Attacks present in the literature.

\subsection{Backdoor Attacks}

A Backdoor Attack in the context of Machine Learning involves an adversary embedding a hidden mechanism within a model during the training process. This mechanism allows the attacker to manipulate the model's behavior in specific ways when triggered by particular input patterns, which are often inconspicuous or seemingly benign. A Backdoor Attack introduces a malicious perturbation $\Delta$, causing misclassification only when the trigger is present while maintaining normal behavior otherwise~\cite{liu2022backdoor}:

\begin{equation}
    F(X) = y_i, \quad \forall X \sim \mathcal{D}_{\text{clean}}
\end{equation}

\begin{equation}
    F(X + \Delta) = y_t, \quad \forall X \sim \mathcal{D}_{\text{backdoor}}
\end{equation}

\noindent
where $X$ represents the clean input data, and $\Delta$ is the trigger pattern introduced by the attacker. The true label of the clean input is denoted as $y_i$, while $y_t$ represents the attacker's target label. The clean data is drawn from the distribution $\mathcal{D}_{\text{clean}}$, whereas the manipulated (backdoored) data comes from the distribution $\mathcal{D}_{\text{backdoor}}$. This definition highlights the stealthy nature of Backdoor Attacks. The model functions normally on clean data, correctly classifying it as $y_i$. However, when the adversarial trigger $\Delta$ is added to the input, the model misclassifies it as the target label $y_t$, allowing the attacker to manipulate predictions without affecting the overall performance on clean samples.

A typical Backdoor Attack consists of the following steps:

\begin{enumerate}
    \item \textbf{Poisoning Phase.} The attacker injects poisoned samples $(X + \Delta, y_t)$ into the training dataset. These samples contain a specific trigger $\Delta$ and are labeled as the target class $y_t$. The goal is to implant the backdoor in the model during training.
    
    \item \textbf{Model Training.} The poisoned dataset, which includes both clean and manipulated samples, is used to train the model. The model learns to classify clean samples correctly but also associates the trigger $\Delta$ with the target label $y_t$.
    
    \item \textbf{Deployment.} The trained model is deployed and behaves normally on clean data, classifying inputs correctly as $y_i$.
    
    \item \textbf{Attack Execution.} When an attacker provides an input $X + \Delta$ containing the trigger, the model misclassifies it as the target label $y_t$. This allows the attacker to control the model’s behavior on specific inputs without significantly affecting overall accuracy.
\end{enumerate}

Backdoor Attacks are a specific subclass of Data Poisoning Attacks where the attacker implants a hidden pattern (trigger) into a subset of training samples and assigns them a target label. In contrast, Availability Attacks (also known as Indiscriminate Poisoning Attacks) aim to degrade the model's overall performance rather than targeting specific inputs. These attacks modify the training data in a way that reduces accuracy across all test samples, causing widespread failure instead of targeted misclassification~\cite{marchant2022hard}. Unlike Backdoor Attacks, availability attacks do not rely on hidden triggers but instead corrupt the model's decision boundaries, making it unreliable.


Current Backdoor defense strategies are broadly classified into backdoor detection and backdoor erasing~\cite{liu2022backdoor}. Detection methods identify whether a model or dataset contains backdoors but do not neutralize them. Erasing techniques aim to mitigate backdoor effects while preserving model accuracy. Fine-tuning with clean data offers a straightforward but weak defense, whereas fine-pruning removes trigger-activated neurons at the cost of performance degradation. More advanced approaches, such as Knowledge Distillation, seek to transfer clean model behavior to the compromised model, offering a more effective defense mechanism.

\begin{figure}
\centering
\scalebox{.7}{
\begin{tikzpicture}[node distance=0.2\linewidth]
\node (start) [box,fill={rgb:black,1;white,5}] {MU and Backdoor Attack};

\node (n0) [box, right of=start, xshift=0.1\linewidth, yshift=0.4\linewidth] {Attack to MU};
\node (n1) [box, right of=start, xshift=0.1\linewidth, yshift=0.15\linewidth] {Defense based on MU};
\node (n2) [box, right of=start, xshift=0.1\linewidth, yshift=-0.15\linewidth] {MU Framework Evaluation};
\node (n3) [box, right of=start, xshift=0.1\linewidth, yshift=-0.4\linewidth] {MU Verification Framework Evaluation \cite{sommer2022athena,guo2023verifying}};

\node (n00) [box, right of=n0,xshift=0.1\linewidth, yshift=0.25\linewidth] {CMU \cite{marchant2022hard,huang2024uba,liu2024backdoor}};
\node (n01) [box, right of=n0,xshift=0.1\linewidth, yshift=0.05\linewidth] {FU \cite{chen2025fedmua}};

\node (n10) [box, right of=n1, xshift=0.1\linewidth, yshift=0.1\linewidth] {CMU \cite{liu2022backdoor,li2023reconstructive,wei2023shared,li2024partially,zhao2024facilitating,niu2024towards}};
\node (n11) [box, right of=n1, xshift=0.1\linewidth,yshift=-0.1\linewidth] {FU \cite{daluwatta2024uaas,wu2024unlearning}};

\node (n20) [box, right of=n2,  xshift=0.1\linewidth, yshift=0.05\linewidth]{CMU \cite{jia2023model,ma2022learn,chen2024private}};
\node (n21) [box, right of=n2, xshift=0.1\linewidth,yshift=-0.1\linewidth]{FU \cite{han2025vertical,jiang2024efficient}};
\draw [arrow] (start) -- (n0);
\draw [arrow] (start) -- (n1);
\draw [arrow] (start) -- (n2);
\draw [arrow] (start) -- (n3);

\draw [arrow] (n0) -- (n00);
\draw [arrow] (n0) -- (n01);
\draw [arrow] (n1) -- (n10);
\draw [arrow] (n1) -- (n11);
\draw [arrow] (n2) -- (n20);
\draw [arrow] (n2) -- (n21);
\end{tikzpicture}}
\caption{Categorization for papers dealing with MU and Backdoor Attacks} \label{fig:backdoorScheme}
\end{figure}
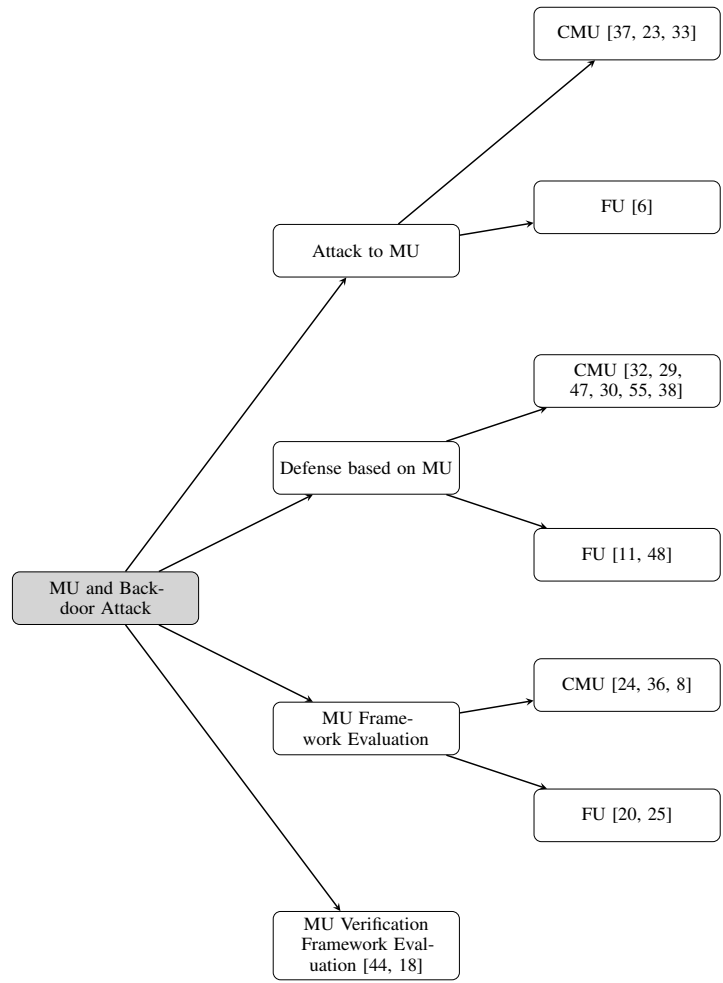

After reviewing the literature related to Machine Unlearning and Backdoor Attacks, we find the following categorization based on the scope of the work, also visible in Figure \ref{fig:backdoorScheme}:

\begin{itemize}
    \item Backdoor Attack against MU \cite{marchant2022hard,huang2024uba,liu2024backdoor,chen2025fedmua};
    \item MU as a defense (or erasing strategy) against Backdoor Attack \cite{liu2022backdoor,li2023reconstructive,wei2023shared,daluwatta2024uaas,zhao2024facilitating,niu2024towards};
    \item Backdoor Attack as an evaluation tool to test new MU framework \cite{jia2023model,ma2022learn,han2025vertical,jiang2024efficient};
    \item Backdoor Attack as a tool for new MU verification framework \cite{sommer2022athena,guo2023verifying,han2025vertical}.
\end{itemize}

Table \ref{tab:Backdoor} illustrates the papers dealing with MU and Backdoor Attacks specifying the unlearning paradigm used between Central Machine Unlearning (CMU) and Federated Unlearning (FU); if the proposed unlearning technique concerns the instance or the class; the attacker knowledge (black-box, gray-box or white-box); the attack/defense phase (during the training or before, during or post the unlearning); the employed defense techniques, and the evaluation metrics used.

\begin{table*}
\centering
\caption{Papers dealing with MU an Backdoor Attack\label{tab:Backdoor}}
\resizebox{\textwidth}{!}{
\renewcommand{\arraystretch}{1.5}
\begin{tabular}{|l|c|p{1.7cm}|p{1.7cm}|p{1.8cm}|p{1.5cm}|p{1.7cm}|p{5cm}|p{5.5cm}|}
\hline
    \textbf{Ref.} & \textbf{Year} & \textbf{Unlearning Paradigm} & \textbf{Class/ Instance MU} & \textbf{Attacker Knowledge} & \textbf{Attack Phase} & \textbf{Defense Phase} & \textbf{Defense Technique} & \textbf{Evaluation Metrics}\\

    \hline
    \cite{liu2022backdoor} & 2022 & CMU & Instance & Black-box & Training & DU & Generative Networks and MU & Attack Success Rate (ASR) and model Accuracy (Acc)\\
    
    \cite{ma2022learn} & 2022 & CMU & Instance & Black-box & PU & BU & Neuron masking & Forgetting Rate\\
    \cite{sommer2022athena} & 2022 & CMU & Instance & Black-box & BU/PU & - & - & Backdoor Attack Success Rate (BASR)\\
    \cite{guo2023verifying} & 2023 & CMU & Instance & Black-box & BU/PU & - & - & Backdoor Attack Success Rate (BASR), Clean Sample Accuracy (CSA)\\
    \cite{jia2023model} & 2023 & CMU & Instance & Black-box & PU & BU & Model sparsification via weight pruning & Unlearning accuracy (UA), MIA-Efficacy, Remaining accuracy (RA), Testing accuracy (TA), Run-time efficiency (RTE)\\
    \cite{li2023reconstructive} & 2023 & CMU & Instance & Black-box & BU & DU/PU & Reconstructive Neuron Pruning (RNP) & Detection Rate (DR), Attack Success Rate (ASR), Clean Accuracy (CA)\\
    \cite{wei2023shared} & 2023 & CMU & Instance & Black-box & BU & DU/PU & Adversarial Training Techniques & Accuracy on benign data (ACC), Attack Success Rate (ASR), Robust Accuracy (R-ACC), Defense Effectiveness Rating (DER)\\
    \cite{daluwatta2024uaas} & 2024 & FU & Instance & Black-box& BU & BU/DU/PU& Gradient Ascent & Expected Calibration Error (ECE), Accuracy, Backdoor Accuracy\\
    \cite{chen2024private} & 2024 & CMU & Instance & Grey-box & PU & DU & Unlearning by small weight perturbation & MIA Evaluation, Backdoor Evaluation, Unlearning Accuracy, Unlearning Time Cost\\
    \cite{huang2024uba} & 2024 & CMU & Instance & Black-box & BU/DU & BU/DU/PU & Outlier Filters, Model Scanners, Anomaly Detectors, Model Reconstructors & Backdoor Effectiveness, Stealthiness, Persistence, Resistance to Defenses\\
    \cite{jiang2024efficient} & 2024 & FU & Class & White-box& DU & PU & Adaptive Differential Privacy (ADP), Dual-layered Selection, Unlearning &Test Accuracy (TA), Membership Inference Success Rate (MISR), Attack Success Rate (ASR)\\
    \cite{li2024partially} & 2024 & CMU &  Instance & Black-box & DU & PU & Partial Training (PT), Super Finetuning (SFT), Neural Attention Distillation (NAD), Anti-Backdoor-Learning (ABL), Data Augmentation & Attack Success Ratio (ASR), Accuracy (Acc)\\
    \cite{liu2024backdoor} & 2024 & CMU & Instance & White-box/ Black-box& DU & DU/PU & Monitoring Unlearning Requests, Model Review and Retraining & Attack Success Rate (ASR), Benign Accuracy (BA), Unlearning Percentage (UP) \\
    \cite{zhao2024facilitating} & 2024 & CMU & Class & White-box/Gray-box & DU & BU/DU/PU & Unlearning-based Model Ablation (UMA) & Detection Accuracy, AUC-ROC, False Positive Rate (FP), False Negative Rate (FN)\\
    \cite{niu2024towards} & 2024 & CMU & Class & White-box & DU & DU/PU & Progressive Unified Defense (PUD), Model Repairing Techniques, Data Filtering, Adversarial Training & Attack Success Rate (ASR), Clean Accuracy\\
    \cite{chen2025fedmua} &2025 & FU & Instance & Black-box & DU  & DU/PU & Gradient Value Adjustments & Attack Success Rate (ASR), Unlearned Global Test Accuracy (gAcc\_G), Clean Global Test Accuracy (Acc\_G)\\
    \cite{wu2024unlearning} & 2024 & FU & Class & White-box & BU & BU & Knowledge Distillation, Purifying Backdoored Models& Test Accuracy (TA), Backdoor Attack Accuracy Rate (AA) \\
    \cite{han2025vertical} & 2025 & FU & Instance & Black-box & BU & DU & Backdoor Certification & Backdoor Accuracy, Kullback-Leibler (KL) Divergence and $\ell$2-Distance, Clean Accuracy\\
    \hline
\end{tabular}
}
\footnotesize BU: Before Unlearning, DU: During Unlearning; PU: Post Unlearning. 
\end{table*}

\subsubsection{Backdoor Attack against MU}

The authors of \cite{marchant2022hard} explore the interplay between MU and Backdoor Attacks, highlighting a novel vulnerability where malicious actors can exploit the unlearning process to embed persistent triggers within datasets. These Backdoor Attacks are strategically designed to maintain model functionality post-unlearning, undermining the efficacy of data erasure mechanisms. By manipulating training data prior to unlearning requests, attackers can create conditions where specific inputs elicit targeted responses, thereby maintaining influence over the model's decisions. This intertwining of backdoor tactics with MU emphasizes the need for robust defenses in Machine Learning systems, as traditional unlearning methodologies may inadvertently facilitate adversarial manipulation. 
 The authors of~\cite{liu2024backdoor} proposed two attack strategies: one implants a backdoor by selectively requesting data removal without modifying training data, while the other injects poisoned samples during training and later activates the backdoor via unlearning requests. To optimize the attack, an objective function selects the unlearning subset and triggers to enhance attack utility while minimizing data removal. 
The discrete unlearning instances are approximated using a differentiable sigmoid function, and the optimization is solved with gradient-based methods. Experiments across various models and datasets demonstrate high attack success with minimal data removal, ensuring efficiency and stealth. 
\par A recent Backdoor Attack, UBA-Inf~\cite{huang2024uba}, exploits MU in Machine-Learning-as-a-Service (MLaaS) to stealthily activate backdoors. Unlike methods proposed in~\cite{liu2024backdoor}, UBA-Inf leverages unlearning requests to remove camouflage samples, prolonging backdoor persistence and evading detection. The attack unfolds in four stages: generating camouflage and backdoor samples, injecting them into training, activating the backdoor through unlearning, and exploiting the model via queries. 
An influence-driven camouflage generation algorithm enhances stealth. UBA-Inf remains effective in both on-demand and continual training MLaaS, highlighting the urgent need for stronger defenses against unlearning-enabled Backdoor Attacks. 
The study~\cite{chen2025fedmua} introduces a novel Backdoor Attack framework termed FedMUA, which exploits the Federated Unearning process to manipulate model predictions maliciously. By strategically initiating unlearning requests aimed at influential training samples, attackers can intentionally misclassify target users while maintaining the accuracy of predictions for non-target users. This stealthy approach raises significant ethical concerns, particularly in sensitive applications like credit scoring, where it can adversely affect individuals' financial reputations. 

\subsubsection{Backdoor Defense based on MU}
MU has been used as a defense mechanism against Backdoor Attack \cite{liu2022backdoor,li2023reconstructive,wei2023shared,daluwatta2024uaas,li2024partially,niu2024towards}.
In \cite{liu2022backdoor}, the authors present BAERASER, a framework that can erase the trigger patterns of a Backdoor Attack from the victim model based on MU leveraging a gradient ascent-based method.
The authors of \cite{li2023reconstructive} propose a defense called Reconstructive Neuron Pruning (RNP) to expose
and prune backdoor neurons via MU. It proceeds firstly by unlearning the neurons on a few clean samples via
a neuron-level unlearning and then recovering the neurons
on the same clean samples via a filter-level recovery. Wei et al. \cite{wei2023shared} propose a framework called  Shared Adversarial Unlearning (SAU) to identify shared adversarial examples and unlearn
them to break the connection between the poisoned sample and the target label.

Also \cite{li2024partially} proposes a defense against Backdoor attacks performing unlearning. In particular, the authors provide a new model training method, called Partial Training(PT), that freezes part of the model to isolate suspicious samples. Zhao et al. \cite{zhao2024facilitating} design UMA a framework able to filter out
backdoor-irrelevant features by unlearning the inherent features of the target class within the model and subsequently
reveals the backdoor through dynamic trigger optimization.
The proposal of \cite{niu2024towards} starts revealing that there is a connection between Backdoor and Adversarial attacks, and then it presents a Progressive Unified Defense (PUD) to jointly erase through unlearning backdoors and enhance the model's adversarial robustness.

The proposals of \cite{daluwatta2024uaas,wu2024unlearning} lie in the context of Federated Unlearning (FU). The authors propose a system that enables
the selective erasure of specific clients' data influence
from the global model in an FL system. Additionally, it provides a solution for cleaning compromised global models by
selectively removing the influence of poisonous clients
without necessitating complete retraining. The authors of \cite{wu2024unlearning} leverage a method able to distinguish the attacker's influence on the
global model by subtracting its historical parameter updates
from the model. To do so, they exploit the Knowledge Distillation method to
remedy the skew caused by the subtraction and do not transfer
any backdoor behaviors.

\subsubsection{Evaluating MU framework through Backdoor-based metrics}
Several works focus on proposing a new MU paradigm considering common attacks to MU to
demonstrate the unlearning efficacy and the robustness of their proposed method \cite{jia2023model,ma2022learn,chen2024private}.
In particular, the work of Jia et al. \cite{jia2023model} proposes a new MU method that leverages model sparsity (achieved by weight pruning) to reduce the gap between approximate unlearning and exact unlearning significantly. In the experiments, the authors present an application of MU to remove the influence of poisoned backdoor data from a learned
model. To do this they hypothesize that an adversary can manipulate a small portion of training data by injecting a
backdoor trigger and modifying data labels towards
a targeted incorrect label. This implies that if the trigger is present at testing there will be an incorrect prediction. They demonstrate that with an appropriate level of sparsity, their method can effectively remove the backdoor effect while largely preserving the model's generalization. Forsaken, the method proposed by Ma et al. \cite{ma2022learn} is based on neuron masking. To further validate the robustness of Forsaken, the authors test it with poisoned data. 
Chen et al. \cite{chen2024private} leverage membership
Inference and backdoor evaluation are used to assess the success of our MU approach, which is based on the introduction of a small perturbation to the model's weights.
Similarly, Han et al. \cite{han2025vertical,jiang2024efficient} lie in the context of FU. In particular, \cite{jiang2024efficient} describes a FU method but it leverages historical information and DP to enhance privacy protection. Whereas, Han et al. \cite{han2025vertical} propose a novel method for Vertical Federated Unlearning and leverage Backdoor Attack to verify the robustness of the unlearning.

\subsubsection{Backdoor in MU Verification frameworks}
Differently from the above approaches, the proposals \cite{sommer2022athena,guo2023verifying} leverage Backdoor
attacks for designing a decentralized MU verification framework in a Machine-Learning-as-a-Service context (MLaaS). In these systems, users can employ prediction results to determine whether providers comply with data deletion requests. In particular, users poison their training samples with unique backdoor triggers linked to target labels before submitting them for model training. Later, they request data deletion and test their backdoor trigger. If the backdoor effect disappears, it confirms deletion; if not, the provider has failed to unlearn the data.

\subsection{Membership Inference Attack}

Membership Inference Attacks (MIAs) threaten training data privacy in Machine Unlearning \cite{carlini2022membership}. In the ML context, MIAs determine whether a data sample was in a model's training set \cite{shokri2017membership,hu2022membership}. The attack follows a security game framework between a challenger and an adversary \cite{carlini2022membership}. Given a trained model \( f_{\theta} \) and a sample \( (x, y) \), the adversary’s membership function is:

\begin{equation}
A(x, y) = \mathds{1}[A_0(x, y) > \tau]
\end{equation}

where \( A_0(x, y) \) is a confidence score and \( \tau \) a decision threshold. A Loss-based MIA infers membership using the loss function:

\begin{equation}
A_{\text{loss}}(x, y) = \mathds{1}[-\ell(f(x), y) > \tau]
\end{equation}

where \( \ell(f(x), y) \) is the model loss, and \( \mathds{1} \) is the indicator function.

Schematically, the attack unfolds through the following step-by-step process.

\begin{enumerate}
    \item \textbf{Train Model.} The challenger trains \( f_{\theta} \) on dataset \( D \sim \mathcal{D} \).
    
    \item \textbf{Sample Data.} A bit \( b \) determines if \( (x, y) \) is from \( D \) (\( b=1 \)) or from \( \mathcal{D} \setminus D \) (\( b=0 \)).
    
    \item \textbf{Send Sample.} The challenger provides \( (x, y) \) to the adversary.
    
    \item \textbf{Adversary Queries Model.} The adversary accesses \( f_{\theta} \) and possibly \( \mathcal{D} \) (e.g., for shadow models).
    
    \item \textbf{Compute Membership Score.} The adversary applies a decision rule, such as loss-based inference.
    
    \item \textbf{Predict Membership.} The adversary classifies \( (x, y) \) as a member or non-member.
    
    \item \textbf{Evaluate Success.} The attack succeeds if the adversary correctly infers \( b \).
\end{enumerate}

Common defense strategies against Membership Inference Attacks (MIAs) include Differential Privacy, which adds noise to training to obscure individual contributions, and regularization techniques (e.g., dropout, weight decay) to reduce model overfitting. Adversarial training enhances robustness by explicitly training against inference attacks, while confidence masking limits the information exposed by model outputs. Knowledge distillation transfers knowledge to a smaller model to remove overfitting artifacts, reducing susceptibility to MIAs. In MU, they can be exploited by an attacker to check if a user's data has truly been removed. MIAs exploit the observation that Machine Learning models often exhibit different behaviors on data they have been trained on compared to unseen data.

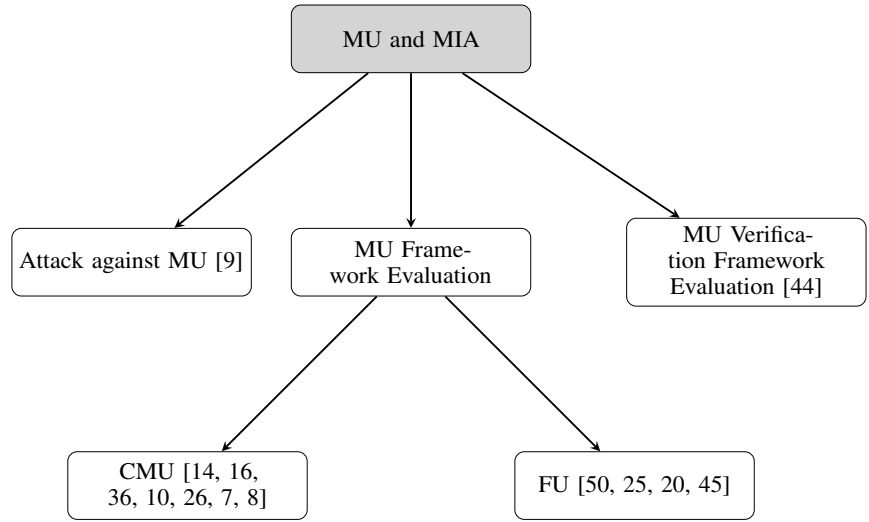
\begin{figure}
\centering
\scalebox{.9}{
\begin{tikzpicture}[node distance=0.2\linewidth]
\node (start) [box,fill={rgb:black,1;white,5}] {MU and MIA};

\node (n5) [box, below of=start,xshift=-0.25\linewidth] {Attack against MU \cite{chen2021machine}};

\node (n1) [box, below of=start] {MU Framework Evaluation};
\node (n2) [box, below of=start,xshift=0.3\linewidth] {MU Verification Framework Evaluation \cite{sommer2022athena}};

\node (n3) [box, below of=n1, xshift=0.2\linewidth] {FU \cite{zhang2023fedrecovery,jiang2024efficient,han2025vertical,varshney2025efficient}};
\node (n4) [box, below of=n1, xshift=-0.2\linewidth] {CMU \cite{golatkar2021mixed,graves2021amnesiac,ma2022learn,chundawat2023zero,kurmanji2023towards,chen2024privacy,chen2024private}};

\draw [arrow] (start) -- (n1);
\draw [arrow] (start) -- (n2);
\draw [arrow] (start) -- (n5);

\draw [arrow] (n1) -- (n3);
\draw [arrow] (n1) -- (n4);

\end{tikzpicture}}
\caption{Categorization for papers dealing with MU and Membership Inference Attacks} \label{fig:MIAScheme}
\end{figure}

After analyzing the literature on Machine Unlearning and Membership Inference Attacks (MIA), we noticed that the works fall into the following three categories:

\begin{itemize}
    \item Membership Inference Attack against MU \cite{chen2021machine};
    \item MIA as evaluation metrics for new MU methods \cite{golatkar2021mixed,graves2021amnesiac,ma2022learn,chundawat2023zero,kurmanji2023towards,chen2024privacy,chen2024private,zhang2023fedrecovery,jiang2024efficient,han2025vertical,varshney2025efficient};
    \item MIA as evaluation metrics for new MU verification methods \cite{sommer2022athena}.
\end{itemize}
 The categorization we employ is visible in Figure \ref{fig:MIAScheme}

Table \ref{tab:MIA} presents the works dealing with MU and Membership Inference Attacks specifying the unlearning paradigm used between Central Machine Unlearning (CMU) and Federated Unlearning (FU); if the proposed unlearning technique concerns the instance or the class; the attacker knowledge (black-box, gray-box or white-box); the attack/defense phase (during the training or before, during or post the unlearning); the employed defense techniques, and the evaluation metrics used.

\begin{table*}
\small
\centering
\caption{Papers dealing with MU an MIA \label{tab:MIA}}
\resizebox{\textwidth}{!}{
\renewcommand{\arraystretch}{1.5}
\begin{tabular}{|l|c|p{1.7cm}|p{1.8cm}|p{2cm}|p{2cm}|p{5cm}|p{5cm}|}
\hline
    \textbf{Ref.} & \textbf{Year} & \textbf{Class/ Instance MU} & \textbf{Attacker Knowledge} & \textbf{Attack Phase} & \textbf{Defense Phase} & \textbf{Defense Technique} & \textbf{Evaluation Metrics}\\

    \hline
    \cite{chen2021machine} & 2021 & Instance & Black-box & PU & DU/PU & Reduce publishing information, Temperature Scaling, Differential Privacy & Degradation Count, Degradation Rate, Attack AUC\\
    
    \cite{golatkar2021mixed} & 2021 & Instance & White-box/ Grey-box & BU/PU & DU/PU & Linear Approximation, Controlled Information Disclosure, Careful Weight Management, Sequential Forgetting Requests & Forgetting Accuracy, Test Time Accuracy, Empirical Risk Minimization\\
    \cite{graves2021amnesiac} & 2021 & Instance & White-box & DU/PU & DU/PU & Unlearning, Amnesiac Unlearning & Model Accuracy, Performance over MIA and Model Inversion Attack\\
    \cite{ma2022learn} & 2022 & Instance & Black-box/White-box & PU & DU/PU & Neuron Masking, Mask Gradient Generator, Dynamic Adjustment of Gradients, Feedback Mechanisms  & Forgetting Rate, Accuracy Loss\\
    \cite{sommer2022athena} & 2022 & Instance & White-box & BU & DU/PU & Neural Cleanse, Neural Attention Distillation & Power of the Hypothesis Test, Detection Accuracy, Impact of User Participation\\
    \cite{chundawat2023zero} & 2023 & Class & White-box & PU & PU & Reducing Model Overfitting, Perturbation of Posteriors, Adversarial Training & Accuracy on Forget Set, Accuracy on Retain Set, Anamnesis Index\\
    \cite{kurmanji2023towards} & 2023 & Instance & Black-box & PU & PU & SCRUB Method, LiRA-for-Unlearning Attack Adaptation, Rewind Strategy & Forget Quality,User Privacy, Utility Metrics, Trade-offs\\
    \cite{chen2024privacy} & 2024 & Class & Grey-box & PU & BU & Adversarial Unlearning, GAN based Unlearning & False Negative Rate, Classification Accuracy, Time Cost\\
    \cite{chen2024private} & 2024 & Instance & Grey-box & PU & DU & Unlearning by small weight perturbation  & MIA Evaluation, Backdoor Evaluation\\
    \cite{zhang2023fedrecovery} & 2023 & Instance & Black-box & DU & PU & Local Differential Privacy, Differentially Private Noise Injection, MIA Strategy  & Accuracy, Running Time, MIA Precision\\
    \cite{jiang2024efficient} & 2024 & Instance & Black-box & DU & PU & Adaptive Differential Privacy, Unlearning Strategies  & Test Accuracy, MIA Success Rate, ASR\\
    \cite{han2025vertical} & 2025 & Instance & Grey-box & PU & DU & Use of Backdoor Triggers, MIA, Constrained Gradient Ascent  & Clean Accuracy,  MIA Performance, Kullback-Leibler (KL) Divergence and $\ell$2-Distance\\ 
    \cite{varshney2025efficient} & 2025 & Instance & White-box & DU & BU & Proof-of-Deniability, Perturbation of Client Updates, Integral Privacy Model, Differential Privacy  & Utility Comparison,  Memory Usage, Retraining Time, Differential Privacy Parameters\\

    \hline
    
\end{tabular}
}
\footnotesize BU: Before Unlearning, DU: During Unlearning; PU: Post Unlearning;  
\end{table*}

\subsubsection{Membership Inference Attack against MU}

The study~\cite{chen2021machine} investigates the vulnerabilities associated with Machine Unlearning (MU) in relation to Membership Inference Attacks (MIA). It emphasizes that while MU aims to delete specific data samples and their influence from a Machine Learning model to protect privacy, it can inadvertently create privacy risks due to the distinct versions of the model—the original and the unlearned—leading to potential information leaks. The authors propose a novel MIA specifically designed for the MU context, which determines if a target sample was part of the training data for the original model. The attack process unfolds in three steps: generating posteriors from both models, constructing a feature vector from these outputs, and then using an attack model to classify the target sample's membership status. Empirical results show that the MIA achieves higher accuracy than classical approaches, highlighting how MU can unintentionally weaken membership privacy.

\subsubsection{Evaluating MU framework through MIA-based metrics}
The papers described in this section leverage a Membership Inference Attack as a MU performance metric demonstrating that the measured inference attack probability is lower in the unlearned model in comparison to the original model for the deleted data. 

Golatkar et al.~\cite{golatkar2021mixed} introduced a mixed unlearning approach leveraging linear approximations and careful weight management. In the experimental campaign, they construct a simple membership attack similar to what they have done in \cite{golatkar2020forgetting} leveraging the entropy of the model output to demonstrate that their forgetting procedure has quite the same attack success as a re-trained model. Similarly, the proposals of \cite{graves2021amnesiac,chundawat2023zero} test their unlearning method performing MIAs against models before and after data removal to show the effectiveness of data removal methods against record-level data leaking. They also show the robustness of their method against Model Inversion Attacks. The authors of \cite{ma2022learn,kurmanji2023towards} propose a metric to measure the effectiveness of a MU
based on the concept of membership inference. In particular, they design the {\em forgetting rate} that describes the transformation rate of the deleted data from
memorized to unknown stages after the unlearning phase. This metric is used to test the proposed dynamic neuron masking approach, called Forsaken in \cite{ma2022learn} and by
Kurmanji et al.~\cite{kurmanji2023towards} in SCRUB, a LiRA-adapted unlearning method to enhance privacy and utility.  Analogously, Chen et al.~\cite{chen2024privacy,chen2024private} use MIA to evaluate
whether the dataset is successfully forgotten or not. 
To execute MIA on the forgotten model, they employ a shadow model training strategy to infer the data and
construct the attack classifier.

Several proposals have been made in the context of Federated Unlearning and performed MIA on the unlearned model to see if the influence of the targeted client was really removed by the
proposed unlearning algorithm \cite{zhang2023fedrecovery,jiang2024efficient,han2025vertical,varshney2025efficient}.
In particular, Zhang et al.~\cite{zhang2023fedrecovery} incorporated local Differential Privacy and noise injection to reproduce a model that is
indistinguishable from the retrained one by only
exploring clients’ historical submissions. Jiang et al.~\cite{jiang2024efficient} define Membership Inference Success Rate
(MISR) to evaluate the FU effectiveness. Han et al~\cite{han2025vertical} studied constrained gradient ascent and backdoor triggers in Vertical FU frameworks validating its effectiveness through MIA.

\subsubsection{Evaluating MU Verification framework through MIA-based metrics}

Sommer et al.~\cite{sommer2022athena} propose a mechanism in the MLaaS context that allows users to verify if the
service provider is compliant with their right to be forgotten. To validate their backdoor-based verification mechanism, they provide the verification performance
by user-level MIAs. In particular, they perform MIA
on each data sample by comparing the prediction confidence to a threshold.

\subsection{Adversarial Attack}

An Adversarial Attack is a deliberate modification of an input to cause an ML model to make an incorrect prediction while keeping the input visually or semantically similar to the original. This modified input is known as an adversarial example. According to Niu et. al \cite{niu2024towards} Adversarial attacks can be classified into targeted and untargeted adversarial attacks. An untargeted adversarial attack seeks to generate a perturbation $r$ that causes an input $x' = x + r$ to be misclassified by an ML model. The objective is to maximize the model's loss $L(x,y)$ with respect to $r$, making the model’s prediction different from the correct label $y$.

$$\max_{r} \mathcal{L}(x', y; \theta), 
\textnormal{subject to} \quad |r|_p \leq \epsilon,$$
$$\quad x' = x + r,
\quad x' \in [0,1]^d$$

\noindent
where $\quad |r|_p$ represents the perturbation constraint under an $l_p$ norm ensuring that $x'$ remains within valid input bounds. An untargeted adversarial attack aims to generate perturbed inputs $x'$ that lead to misclassification, meaning the model's prediction differs from the true label $y$. Unlike a targeted adversarial attack, which forces the model to classify $x'$ as a specific target label, an untargeted attack only seeks to disrupt the original classification without dictating the incorrect class. Consequently, research has shown that the predicted labels of adversarial examples tend to follow a uniform distribution across all possible classes. 

While classical defenses in Machine Learning have traditionally focused on building robustness directly into the model's architecture or training process, Machine Unlearning offers a complementary approach. Instead of solely focusing on making the model inherently resilient to perturbations, Machine Unlearning allows for the targeted removal of specific, potentially harmful, learned patterns. This is particularly relevant in the context of adversarial attacks, where the attack's success may rely on the model exploiting specific, learned correlations within the training data. By selectively ``forgetting'' these correlations, Machine Unlearning can effectively neutralize the attack's effectiveness. 

\begin{figure}
\centering
\scalebox{.9}{
\begin{tikzpicture}[node distance=0.2\linewidth]
\node (start) [box,fill={rgb:black,1;white,5}] {MU and Adversarial Attack};

\node (n0) [box, below of=start, xshift=-0.2\linewidth] {Attack to MU \cite{zhao2023static,gupta2021adaptive}};
\node (n1) [box, below of=start, xshift=0.2\linewidth] {Defense based on MU \cite{niu2024towards,liu2023muter}};

\draw [arrow] (start) -- (n0);
\draw [arrow] (start) -- (n1);

\end{tikzpicture}}
\caption{Categorization for papers dealing with MU and Adversarial Attacks} \label{fig:AdvScheme}

\end{figure}
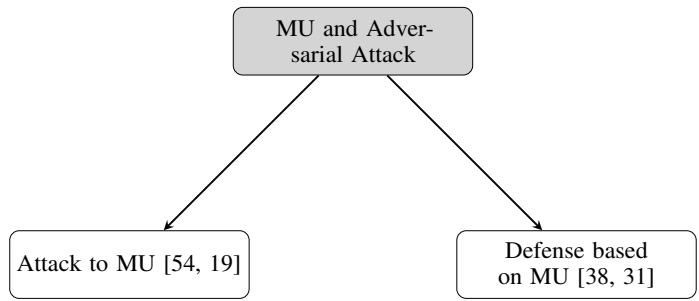

After examining the work on MU and Adversary Attacks, we observed that several works present MU as a tool for the defense against adversary attacks \cite{niu2024towards,liu2023muter}, whereas a contribution proposes an Adversarial Attack to MU \cite{zhao2023static}. The categorization we employ is visible in Figure \ref{fig:AdvScheme}.

Table \ref{tab:adversarial} illustrates the work dealing with MU and Adversarial Attacks specifying the unlearning paradigm used between Central Machine Unlearning (CMU) and Federated Unlearning (FU); if the proposed unlearning technique concerns the instance or the class; the attacker knowledge (black-box, gray-box or white-box); the attack/defense phase (during the training or before, during or post the unlearning); the employed defense techniques, and the evaluation metrics used.

\begin{table*}
\centering
\caption{Papers dealing with MU an Adversarial Attack\label{tab:adversarial}}
\resizebox{\textwidth}{!}{
\renewcommand{\arraystretch}{1.5}
\begin{tabular}{|l|c|p{1.7cm}|p{1.7cm}|p{1.8cm}|p{1.5cm}|p{1.7cm}|p{5cm}|p{5.5cm}|}
\hline
    \textbf{Ref.} & \textbf{Year} & \textbf{Unlearning Paradigm} & \textbf{Class/ Instance MU} & \textbf{Attacker Knowledge} & \textbf{Attack Phase} & \textbf{Defense Phase} & \textbf{Defense Technique} & \textbf{Evaluation Metrics}\\

    \hline
    \cite{gupta2021adaptive} & 2021 & CMU & Instance & White-box & BU/DU & DU & Differential Privacy, Adaptive Unlearning Algorithms, Oblivious Sequence Assumption  & Deletion Guarantee, Indistinguishability, Computational Cost\\    \cite{liu2023muter} & 2023 & CMU & Instance & White-box & BU & BU/DU & Adversarial Training  & Effectiveness, Accuracy, Robustness, Efficiency \\
    \cite{zhao2023static} & 2023 & CMU & Instance & Black-box/ White-box & DU & DU & - & Demographic Parity, Equalized Odds\\
    \cite{niu2024towards} & 2024 & CMU & Instance & White-box & BU/DU & DU/PU & Adversarial Training, Progressive Unified Defense (PUD), Hybrid Approaches & Robust Accuracy, Clean Accuracy, ASR \\

    \hline
\end{tabular}
}
\footnotesize BU: Before Unlearning, DU: During Unlearning; PU: Post Unlearning. 
\end{table*}
\subsubsection{Adversarial Attacks Against Machine Unlearning}

Zhao et al.~\cite{zhao2023static} introduce two attack frameworks that exploit vulnerabilities in Machine Unlearning (MU) systems: static selective forgetting and sequential selective forgetting attacks. In a \textit{static selective forgetting attack}, the adversary submits a batch of malicious data update requests simultaneously to manipulate the unlearning process. This attack operates as follows:
\begin{itemize}
    \item The adversary crafts data updates designed to interfere with selective forgetting, potentially leading to misclassification or exacerbating biases.
    \item The attack employs discrete indication variables to specify deletions, making direct solutions intractable. To overcome this, the authors propose an approximation using continuous differentiable functions.
\end{itemize}

In contrast, a \textit{sequential selective forgetting attack} strategically submits data updates over time to maximize damage while minimizing detection risk:
\begin{itemize}
    \item The adversary carefully times and selects updates to exploit the unlearning process dynamically.
    \item The attack is formulated as a stochastic optimal control problem, focusing on order and timing to achieve adversarial goals.
    \item By selectively modifying, adding, or deleting updates, the adversary influences the system’s behavior progressively.
\end{itemize}

The evaluation considers both white-box and black-box settings across unlearning methods such as first-order, second-order, unrolling SGD, amnesiac, and SISA. The results demonstrate consistently high attack success rates across different datasets and MU techniques. Traditional defenses against adversarial attacks, such as adversarial training, input preprocessing, gradient masking, and defensive distillation, focus on robustness against small perturbations but often fail as attackers develop adaptive strategies. To assess unlearning robustness, Gupta et al.~\cite{gupta2021adaptive} evaluate model privacy through model inversion attacks, demonstrating that Differential Privacy-based unlearning can mitigate adaptive deletion threats by limiting information leakage.

These attack frameworks highlight critical security challenges in MU, emphasizing the need for stronger adversarial resilience in unlearning mechanisms.

\subsubsection{Defense against Adversarial Attack based on MU}
Adversarial Training Models (ATMs) are a defense strategy that enhances model robustness by training on adversarial examples. 
The proposals of \cite{niu2024towards,liu2023muter} describe a defense against Adversarial through MU.
In particular, Niu et al. \cite{niu2024towards} reveal the connection between Backdoor and Adversarial attacks, and then it presents a Progressive Unified Defense (PUD) to jointly erase through unlearning backdoors and enhance the model's adversarial robustness. The authors of \textit{MUter} \cite{liu2023muter} propose a novel MU approach for ATMs, introducing a closed-form unlearning step based on a total Hessian-related data influence measure. This method addresses limitations in existing techniques that struggle to capture the indirect Hessian component of data influence accurately.

\subsection{Inversion Attacks}
Inversion Attacks can be classified in {\em Model Inversion Attacks} and {\em Gradient Inversion Attacks}. 
Model Inversion (MoIAs) attacks exemplify privacy risks in ML. These attacks leverage the correlation between training data and model outputs to reconstruct sensitive attributes of the training data. MI attacks typically formulate this reconstruction as an optimization problem, aiming to find the sensitive feature values that maximize the likelihood given the target model\cite{zhang2020secret}.
MI aims to reconstruct sensitive features \( x_1 \) of an input \( x \) given partial knowledge of the target model \( f \). Formally, let:

\begin{itemize}
    \item \( x = (x_1, x_2, \dots, x_t) \) be the feature vector of an individual.
    \item \( y = f(x) \) be the model's predicted output.
    \item The adversary has access to the model \( f \) and auxiliary information:
    \begin{equation}
        \text{side}(x, y) = (x_2, \dots, x_t, y).
    \end{equation}
\end{itemize}

The attack infers the sensitive feature \( x_1 \) by solving:

\begin{equation}
    \hat{x}_1 = \arg\max_{x_1} P(x_1 \mid x_2, \dots, x_t, y, f).
\end{equation}

For a linear regression model \( f(x) = w_1 x_1 + w_2 x_2 + \dots + w_t x_t + b \), the adversary reconstructs \( x_1 \) as:

\begin{equation}
    \hat{x}_1 = \frac{y - (w_2 x_2 + \dots + w_t x_t + b)}{w_1}.
\end{equation}

Model Inversion Attacks generally proceed through the following sequential phases.
\begin{enumerate}
    \item \textbf{Attack Setup:} The adversary has \textit{white-box access} to the model \( f \) and knows all features except the sensitive one (\( x_1 \)).
    \item \textbf{Inference Process:} Using the model’s parameters and auxiliary data, the adversary reconstructs \( x_1 \).
    \item \textbf{Reconstruction Strategy:} 
    \begin{itemize}
        \item If \( f \) is linear, \( x_1 \) is computed algebraically.
        \item For complex models, optimization or gradient-based methods are used.
    \end{itemize}
\end{enumerate}

Model Inversion Attacks exploit \textit{model transparency} to infer private attributes, posing severe privacy risks in sensitive domains like healthcare and biometrics. While traditional MoIAs focus on pre-unlearning data, the concern here lies in extracting information about unlearned data, exploiting potential residual information left within the model, even after the unlearning process. MoIA's are also used as an evaluation metric as a model's performance over it can prove the efficiency of unlearning.

Gradient Inversion Attacks (GIAs) exploit shared model weights and gradients to leak private data by reconstructing data or labels through minimizing differences between observed and synthetic gradients~\cite{hu2024learn}. Attackers generate dummy gradients using random inputs and iteratively optimize until the reconstructed data reaches minimal error, following either iterative or recursion-based paradigms~\cite{ijcai2022p791}.

Model Inversion Attacks and Gradient Inversion Attacks are both privacy threats that aim to recover sensitive information from Machine Learning models. MoIAs use model outputs and auxiliary knowledge to infer missing attributes, often targeting deployed models with white-box or black-box access. In contrast, GIAs exploit shared gradients in distributed learning settings, such as Federated Learning, to reconstruct entire input samples. While MoIAs typically infer partial information (e.g., sensitive features), GIAs can fully recover training data, making them more severe in collaborative learning environments. Despite their differences, both attacks highlight vulnerabilities in model transparency and training data exposure.

Traditional defense strategies against Inversion Attacks include regularization techniques like dropout and weight decay, which enhance generalization and reduce overfitting to training data. Differential privacy injects controlled noise during training to obscure the direct relationship between inputs and outputs. Adversarial training strengthens models by exposing them to inversion attempts during training, improving robustness. Additionally, Knowledge Distillation transfers learned knowledge to a new model while abstracting sensitive details, mitigating privacy leakage.

After reviewing the literature related to Machine Unlearning and Model Inversion Attacks, we find the following categorization:

\begin{itemize}
    \item Model Inversion Attack to MU \cite{hu2024learn};
    \item Model Inversion Attack as an evaluation metric for MU \cite{graves2021amnesiac,chundawat2023zero};
    \item MU as a defense against Gradient Inversion Attack \cite{gao2024defending}
\end{itemize}

\begin{figure}
\centering
\scalebox{.9}{
\begin{tikzpicture}[node distance=0.2\linewidth]
\node (start) [box,fill={rgb:black,1;white,5}] {MU and Inversion Attack};

\node (n0) [box, below of=start, xshift=-0.2\linewidth] {Model Inversion Attack};
\node (n1) [box, below of=start, xshift=0.2\linewidth] {Gradient Inversion Attack};

\node (n2) [box, below of=n0, xshift=0.2\linewidth] {Attack to MU \cite{hu2024learn}};
\node (n3) [box, below of=n0, xshift=-0.1\linewidth] {MU Framework Evaluation \cite{graves2021amnesiac,chundawat2023zero}};
\node (n4) [box, below of=n1, xshift=0.1\linewidth] {MU Defense based on MU \cite{gao2024defending}};
\draw [arrow] (start) -- (n0);
\draw [arrow] (start) -- (n1);
\draw [arrow] (n0) -- (n2);
\draw [arrow] (n0) -- (n3);
\draw [arrow] (n1) -- (n4);

\end{tikzpicture}}
\caption{Categorization for papers dealing with MU and Inversion Attacks} \label{fig:inversionScheme}
\end{figure}
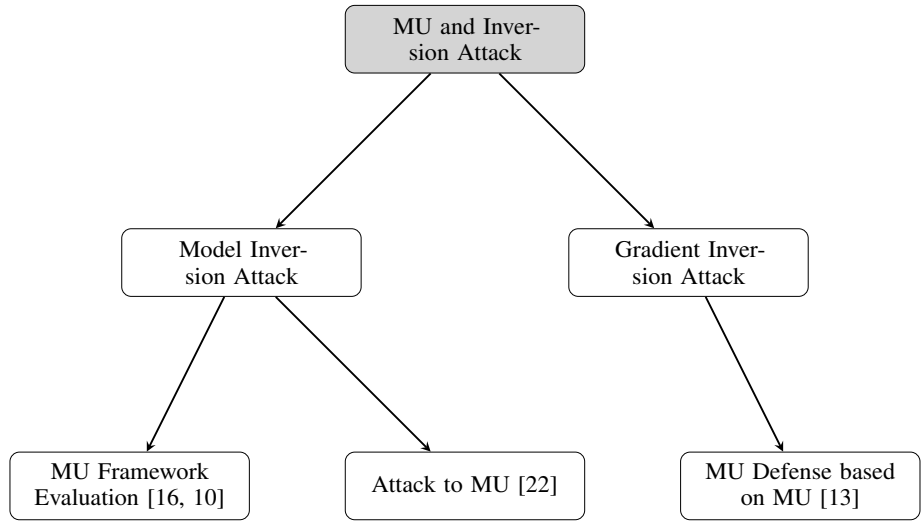

The categorization we employ is visible in Figure \ref{fig:inversionScheme}. Whereas, table \ref{tab:inversion} describes the papers dealing with MU and Model Inversion Attacks specifying the unlearning paradigm used between Central Machine Unlearning (CMU) and Federated Unlearning (FU); if the proposed unlearning technique concerns the instance or the class; the attacker knowledge (black-box, gray-box or white-box); the attack/defense phase (during the training or before, during or post the unlearning); the employed defense techniques, and the evaluation metrics used.

\begin{table*}
\centering
\caption{Papers dealing with MU an Model Inversion Attack\label{tab:inversion}}
\resizebox{\textwidth}{!}{
\renewcommand{\arraystretch}{1.5}
\begin{tabular}{|l|c|p{1.7cm}|p{1.7cm}|p{1.8cm}|p{1.5cm}|p{1.7cm}|p{5cm}|p{5.5cm}|}
\hline
    \textbf{Ref.} & \textbf{Year} & \textbf{Unlearning Paradigm} & \textbf{Class/ Instance MU} & \textbf{Attacker Knowledge} & \textbf{Attack Phase} & \textbf{Defense Phase} & \textbf{Defense Technique} & \textbf{Evaluation Metrics}\\

    \hline
    \cite{graves2021amnesiac} & 2021 & CMU & Instance & White-box & DU/PU & DU/PU & Unlearning, Amnesiac Unlearning & Model Accuracy, Performance over MIA and Model Inversion Attack \\
    \cite{chundawat2023zero} & 2023 & CMU & Class & White-box & PU & PU & Reducing Model Overfitting, Perturbation of Posteriors, Adversarial Training & Accuracy on Forget Set, Accuracy on Retain Set, Anamnesis Index\\
    \cite{gao2024defending} & 2024 & FU & Instance & Gray-box & PU & DU & Knowledge Distillation, Statistical Machine Unlearning, Parameter Cropping, Gradient Perturbation& Model Accuracy, Privacy Protection, Communication Efficiency\\
    \cite{hu2024learn} & 2024 & CMU  & Instance & Black-box/ White-box & PU & DU & Parameter Obfuscation, Model Pruning, Fine-tuning & Mean Squared Error (MSE), Peak Signal-to-Noise Ratio (PSNR), Learned Perceptual Image Patch Similarity (LPIPS)\\
 
    \hline
\end{tabular}
}
\footnotesize BU: Before Unlearning, DU: During Unlearning; PU: Post Unlearning. 
\end{table*}

\subsubsection{Model Inversion Attack against MU}
The study~\cite{hu2024learn} presents unlearning inversion attacks, a novel privacy threat targeting Machine Unlearning in deep neural networks. These attacks reveal sensitive information about unlearned data through two primary methods: feature leakage and label leakage. In the feature leakage scenario, adversaries with white-box access to both original and unlearned models utilize gradient inversion techniques to reconstruct the features of unlearned samples from changes in model parameters. Conversely, in the label leakage scenario, attackers with black-box access generate probing samples to assess discrepancies in predictions between the original and unlearned models, enabling them to infer the class labels of the unlearned data. The paper substantiates these vulnerabilities through extensive experiments, underscoring significant privacy risks associated with Machine Unlearning methodologies.

\subsubsection{Model Inversion Attack as an evaluation metric for MU}
The proposal of Graves et al.~\cite{graves2021amnesiac} and Chundawat et al. \cite{chundawat2023zero} test the robustness of their methods against Model Inversion Attacks. In particular, \cite{graves2021amnesiac} applies a modified version of the standard model inversion
attack \cite{fredrikson2015model} and
defines amnesiac unlearning and considers the case in which the adversary
does not have information about what each class represents. They assume an attack is successful if the adversary is able
to glean information about what the class represents through
model inversion. 
Chundawat et al.\cite{chundawat2023zero} evaluate data leakage in their proposed zero-shot MU method using model inversion attacks. They assume a white-box adversary with access to the unlearned model but not previous versions.  They adapt a known model inversion attack, initializing with noise and optimizing via gradient descent against a target class. Attacks are performed on a fully trained model, a retrained model (without the target class), and their ``forget'' model. Successful attacks are judged by the ability to infer class representations. Their results show successful inversion on the fully trained model, but random patterns on the retrained and forget models, indicate robustness.

\subsubsection{MU as a defense against Gradient inversion attack}

The proposal of \cite{gao2024defending} leverages statistical MU methods in the context of Federated Learning. The main goal of the paper is to present a defense method against gradient inversion attacks inspired by the principles of statistical MU. To do so, it performs a shift from individual data to statistical summaries, creating an abstraction layer to safeguard privacy and facilitate selective data forgetting. The approach disrupts the transformation from raw data to gradients by replacing them with gradients computed from statistical information and employs teacher-student models trained with dual loss functions to confuse adversaries (curious but honest servers).

\section{Challenges and Open Problems}
\label{sec:challenge}

As unlearning techniques evolve and their adoption grows, new challenges and unresolved issues continue to emerge, demanding further investigation. Therefore, the following emerging challenges should be explored as potential research directions that could enhance the MU systems' security, reliability, and effectiveness. 

\begin{itemize}
    \item \textbf{Privacy-preserving MU}. Existing MU systems typically assume that the data slated for removal is directly accessible to the server performing the unlearning process. But, especially in decentralized scenarios such as the one of Federated Unlearning, the service provider should not have access to users' data, raising critical concerns about how to ensure the privacy of the data being unlearned. 
    \item \textbf{MU for Large Models}. MU becomes significantly more complex when dealing with large-scale models, such as deep neural networks (DNNs) and transformer-based architectures (e.g., GPT, BERT). The challenges are related to the scalability, privacy, security, and stem from the size and distributed nature of these models. 
    \item \textbf{Ethical and Regulatory Considerations}. Ensuring that all traces of personal or sensitive information are fully removed is a complex task. Legal standards require verifiable proof of unlearning, which can be hard to guarantee.

    \item \textbf{Certified MU Framework with Blockchain Integration}. Traditional Machine Unlearning lacks transparency, preventing users from verifying data removal. Blockchain integration may offer verifiable, auditable, and tamper-proof unlearning but might introduce challenges like computational overhead and scalability. Future research must optimize efficiency while ensuring security and practical deployment.
    
\end{itemize}

\section{Conclusion}
\label{sec:conclusion}
In the current Machine Learning (ML) context, security threats are becoming more sophisticated, posing serious risks to data privacy and model integrity. 
Recently a new paradigm known as Machine Unlearning (MU) has been designed to enable data removal in compliance with privacy laws (e.g., GDPR's Right to Be Forgotten) and to mitigate security risks.
However, the relationship between ML threats and MU remains underexplored. 
To tackle this issue, this Systematization of Knowledge (SoK) studies current papers dealing with MU techniques and ML security threats to unfold their hidden relationships. To do so, we examined four key attack classes that are the most used in scientific literature, namely Backdoor Attacks, Membership Inference Attacks (MIA), Adversarial Attacks, and Inversion Attacks, analyzing their impact on MU and the role of MU in mitigating them. Additionally, we classified the interaction between these ML threats and MU into three main perspectives: {\em(i)} attacks against MU, {\em(ii)} MU as a defensive mechanism to counteract attacks, {\em(iii)} attacks as evaluation tools to assess the effectiveness of MU frameworks, and {\em(iv)} attacks as verification tools to ensure MU guarantees. Our study identifies critical gaps in existing defenses based on MU and verification methodologies, highlighting avenues for future research. This SoK establishes a foundation for the development of robust, verifiable, and attack-resilient MU solutions in the evolving ML security landscape highlighting key challenges and future research directions in this field.

The research works examined in this paper serve as a foundation for further exploration, as we intend to deepen our investigation into specific aspects that were only briefly discussed in this survey. For example, an interesting research direction is the analysis of existing benchmark datasets for testing MU frameworks. Additionally, exploring insights from legal, ethical, and regulatory perspectives to align MU with data protection laws like GDPR can also be considered as future analysis.
\bibliographystyle{splncs04}
\bibliography{biblio}

\end{document}